\documentclass[pre,amsmath,amssymb,aps,nofootinbib,reprint]{revtex4-2}

\usepackage{url}
\usepackage{amsmath,amssymb,graphicx,bbm}
\usepackage{mwe}
\usepackage{rotating}
\usepackage[caption=false]{subfig}
\usepackage{calc}
\usepackage{upgreek}
\usepackage{circuitikz}

\graphicspath{ {./} }

\usepackage{graphicx}
\usepackage{dcolumn}
\usepackage{bm}
\usepackage{paralist} 
\DeclareMathOperator{\mc}{\enspace ,}

\usepackage[colorlinks,citecolor=blue,linkcolor=blue]{hyperref}
\usepackage[]{cleveref}
\Crefname{figure}{Fig.}{Figs.}
\Crefname{table}{Tab.}{Tabs.}
\Crefname{equation}{Eq.}{Eqs.}

\begin{document}
\preprint{}

\title{Shape and Performance of Fastest Paths over Networks with Interacting Selfish Agents}

\author{Marco Cogoni}
\author{Giovanni Busonera}
\author{Enrico Gobbetti}
\affiliation{CRS4 - Center for Advanced Studies, Research and Development in Sardinia}
\address{Ex Distilleria - Via Ampere 2, 09134 Cagliari (CA) Italy}
\email{\{marco.cogoni, giovanni.busonera, enrico.gobbetti\}@crs4.it}

\date{\today}

\begin{abstract}
We study the evolution of the fastest paths (FP) in transportation networks under increasing congestion. 
Moving from the common edge-based to a path-based analysis, we examine the directed FPs connecting random origin-destination pairs as traffic grows. We describe their shape through effective length, detour (maximum distance of FP from a straight line), inness (signed area between FP and straight line), and their performance through a novel metric measuring how fast and how far an agent travels toward its destination.
The entire network is characterized by analyzing the distribution of the performance metric (and its Gini coefficient) across uniformly sampled paths.
The study focuses on the traffic loading phenomenon that takes place during the morning peak hour for eight major cities: Networks start with empty edges that are progressively populated by the FPs of single vehicles. As vehicle density grows, the interactions among selfish agents becomes stronger at edge level, and travel speed linearly decreases, thus optimal paths dynamically change with traffic.
We fully characterize the transition to congestion and discuss the common aspects among the cities (and some peculiarities), in particular we were able to pinpoint a critical traffic level (or a sequence) for which path shape, rejection ratio, and inequality of the performance degradation, show a concurrent qualitative change. For all cities we observe large peaks for both detour and inness (and their variance) in the proximity of the critical traffic level.
Inness shows that paths are slightly attracted by city centers with light traffic, but switch to a strong repulsion immediately  beyond the transition. Finally, our path performance metric highlighted a strongly asymmetric behavior when the city neighborhoods act as origins or destinations. 

\end{abstract}

\maketitle

\section{Introduction}
\label{sec:intro}

Network transport performance and robustness are usually estimated using static measures depending on topological and geometrical factors that characterize the network in its idle state, but cannot grasp the complex structural modifications that happen when the network is subject to growing levels of traffic~\cite{helbing_traffic_2001,zeng_switch_2019,zhang2022complex}. Network robustness is often studied by removing random elements such as nodes and edges or varying their capacity to highlight weak points and bottlenecks within the system~\cite{Magnien:2021:IRF}. While very useful for evaluating abstract network geometries, this approach does not consider any specific process happening over the network. It turns out, however, that details of the transport process may greatly impact how the system reacts to perturbations. In particular, real processes typically produce non-random, spatially correlated, perturbations, thus knowledge about them allows the development of better testing procedures~\cite{oehlers2021graph,schwarze2024structural}. In this context, percolation theory is sometimes used to model transport disruption~\cite{zeng_switch_2019,zeng_multiple_2020}, and it can be used to estimate the fraction of malfunctioning edges (or nodes) sufficient to break up the network into several disconnected clusters~\cite{cogoni2021stability}. This approach has been recently applied to traffic in urban networks by using the vehicle speed on each road (edge) as a predictor of transport collapse to find the threshold speed above which fragmentation appears for different traffic regimes~\cite{cogoni2024predicting}. Percolation, however, ignores the details of the transport process altogether, modeling only its effects. 

Estimating whole-network performance using the distribution of edge efficiencies may give a biased picture depending on the specific transport process. For example, when \emph{Origins} and \emph{Destinations} (OD) are connected via shortest (or fastest) paths, the impacted edges are spatially correlated and generally produce a strongly inhomogeneous load profile. Different requirements on path routing and OD spatial distribution may lead to entirely different network usage patterns, thus translating to different expected performances. Transport processes are found everywhere in natural and artificial systems, for instance in urban networks, whose traffic patterns have been strongly influenced by the emergence of personal navigation tools that provide the fastest paths (FPs) in real time~\cite{serok_unveiling_2019}.

A non-local measure widely employed to estimate network performance at the edge level is the \emph{Betweenness Centrality} (BC). Centrality measures have long been used to predict which edges (and nodes) are typically subject to the highest traffic, but the correlation with real traffic is known to suffer in the medium and high-density regimes~\cite{holme2003congestion, guimera2005worldwide}. Edge usage obtained from BC, in fact, typically mimics a low-density state of the network that usually happens with very small (compared to network geometry) or very fast (with respect to congestion buildup timescales) agents~\cite{kazerani2009can}. 

Recently, standard BC has been improved by introducing a model to simulate urban networks with different traffic levels and considering the strong interaction effects observed in different conditions to obtain a better measure of edge congestion~\cite{cogoni2024predicting}. While the theoretical foundations of that approach are transferable to other contexts~\cite{yeung2012competition, hamedmoghadam_percolation_2021, olmos_macroscopic_2018, colak_understanding_2016}, urban networks have specific connectivity characteristics that have led to the emergence of specific theoretical models and studies. In particular, they belong to the special class of (almost) planar graphs~\cite{diet_towards_2018,helbing_traffic_2001} whose topology is constrained by the geographical embedding. This severely hinders their long-range connectivity and limits their maximum node degree~\cite{aldous2013true}. The specificity of urban networks has been widely studied~\cite{colak_understanding_2016,li_percolation_2015,kirkley_betweenness_2018,serok_unveiling_2019}, both in terms of their growth over time and of their complex dynamics under different traffic levels.

In this work we propose a shift in the performance measuring process departing from an edge-based in favor of a path-based network metric. The basic idea relies on the fact that having the majority of network components that are still efficient is not enough to guarantee functional transports if the assigned task often implies the use of a minority of saturated links. This is especially crucial for transport processes such as the ones occurring over urban networks, where the congestion patterns are shaped by interacting traveling agents that dynamically react to the growing traffic by selfishly adapting their routing: A behavior that is rapidly becoming dominant within our cities.

Our main goal in this work is to study the evolution of the shape and the performance distribution of the FPs while increasing the traffic level from an empty to a deeply congested network. To this aim, we focus on the scaling behavior of FP effective path length (defined in \ref{subsec:interaction_model}) and detour (maximum distance of FP from a straight line) for a set of large, densely-populated, metropolitan areas. Critical exponents are computed for growing network congestion and plotted on a plane to better understand how they evolve. To better characterize the relation of the FP morphology with its embeddings over a broad range of traffic, we also analyze inness~\cite{lee_morphology_2017} (signed area between FP and straight line) and we discover that the slightly attractive force exerted by city centers (when empty) switches to strong repulsion beyond a well-defined traffic level.
Moving beyond path morphology, we discuss the evolution of the path performance distribution: Path performance turns out to be useful not only in better evaluating whole-network efficiency, but in uncovering a strong asymmetry of different parts of the cities when acting as sources or destinations of traffic. Finally, we resort to the Gini coefficient of the performance distribution to estimate the geographical inhomogeneity of the effects of congestion on path slowdown.

\section{Methods}
\label{sec:methods}

\subsection{Interaction model}
\label{subsec:interaction_model}

In real urban networks, the average travel time is of the same order of magnitude as the typical timescale of congestion buildup, which has been measured to be about one hour in major cities~\cite{li_percolation_2015}. The strength of the interaction among vehicles depends on the local traffic density given by the probability of vehicle coexistence on the same edge during rush hours. This allows for estimating the cumulative traffic seen on the roads during that finite time window. The interaction model used in this work does not incorporate full vehicle dynamics but takes into account the contribution to the traffic of each vehicle added to the network at a road segment level by updating the expected travel times at each step. An earlier work~\cite{cogoni2024predicting} has shown how this model, though simplified, can describe qualitative and quantitative properties of the network loading process of synthetic and real urban networks. We summarize here the aspects of the model relevant to this work and refer the reader to the original publication for further model details~\cite{cogoni2024predicting}. 

Time is not modeled explicitly, but the interval between the empty network and congestion, which represents the rush-hour period, is divided into a sequence of vehicle creation steps, which may be added one at a time or in batches. At each step, the fastest route between O and D is computed with complete knowledge of the network state, and the vehicle density (hence speed) on each affected edge is updated. Vehicles are never removed from the network, even the ones traveling for very short paths, though their impact on the network will be proportional to the ratio of expected travel time on the affected edges and the chosen simulation duration ($\tau$). The relation between current speed and density on a single edge follows the single regime Greenshields model~\cite{helbing_traffic_2001,rakha2002comparison}, for which speed starts as free flow on an empty road, decreasing linearly to zero with maximum density. We use the concept of (delocalized) vehicle interchangeably with that of its path since no position in time exists.

We simulate the network evolution, as observed by travelers, while the traffic $V$ increases from zero up to a target number of vehicles, enough to bring all cities to an almost complete gridlock, signaled by the vanishing probability of completing new paths in finite time. The traffic network is modeled as a directed, weighted graph whose edges represent road segments between adjacent intersections (nodes) and possess three constant features: physical length $l$, speed limit $v^*$, and number of lanes $c$. Nodes, on the other hand, are featureless. Travel time $t$ on each edge is defined as the ratio of its length and current speed $v$. As we compare our results for real cities against synthetic networks~\cite{kartun2019shape} in which speed plays no role, we introduce the concept of effective path length as $l\cdot{v^*}/{v}$, thus encoding speed variations into geometrical deformations. Hereafter we will use travel time and effective path length interchangeably.     

The network traffic grows incrementally by activating one new path at each simulation step, to reach the desired target volume at time $\tau$ (the end of the simulation).
Thus, simulation steps may be interpreted both as the current number of added paths and as a temporal marker within the sequence of OD pairs randomly generated for each simulation. O and D are chosen uniformly over the network nodes. All paths, regardless of their creation order, have the same time constraints.

Intuitively, the occupancy factor $Q_e$ induced by a vehicle over an edge $e$ is the time the vehicle is supposed to spend on it (as a fraction of $\tau$), as forecast at its departure.

The normalized vehicle density is defined by:
\begin{equation}
      \rho_e = \frac{\sum Q_e L}{l_e c_e} \in [0, 1] \mc
    \label{density_def}
\end{equation}
at each simulation step, where $L$ is the average space occupied by one vehicle, and the denominator represents the edge capacity (number of lanes $c_e$ times edge length $l_e$). The sum over $Q_e$ is over all paths using that edge. We call the sum of all edge capacities the maximum vehicular capacity of the whole network.
Since the initial vehicles find a nearly empty network, their FPs
and travel times are virtually equivalent to the non-interacting case. With rising traffic, however, edges fill up and the previous FPs will
disappear and less-used roads and residential neighborhoods will be chosen.
Some edges will eventually reach maximum density and become dysfunctional.
If the fastest route from O to D comprises a dysfunctional edge
(i.e., the network is disconnected) we add just the initial part of the path, skipping all remaining edges from the first dysfunctional one.

\subsection{Path shape analysis}
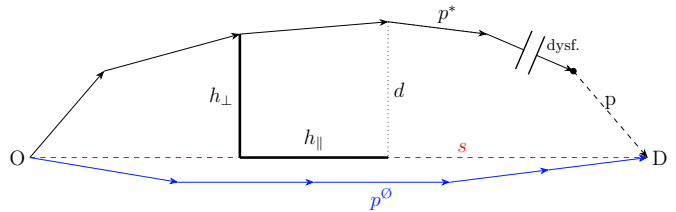
\begin{figure}[htb]
\centering
\resizebox{0.5\textwidth}{!}{%
\begin{circuitikz}
\tikzstyle{every node}=[font=\large]
\draw [ color={rgb,255:red,255; green,0; blue,0}, ->, >=Stealth, dashed] (5.75,7.75) -- (18.25,7.75)node[pos=0.7,above, fill=none]{$s$};
\draw [->, >=Stealth] (5.75,7.75) -- (7.25,9.5);
\draw [->, >=Stealth] (7.25,9.5) -- (10,10.25);
\draw [->, >=Stealth] (10,10.25) -- (13,10.5);
\draw [->, >=Stealth, dotted] (13,10.5) -- (15,10.25)node[pos=0.6,above, fill=none]{$p^*$};
\draw [->, >=Stealth] (13,10.5) -- (15,10.25)
to[capacitor,-*] (16.75,9.5) node[shift={(-0.2,0.5)}] {{\small dysf.}};
\draw [->, >=Stealth, dashed] (16.75,9.5) -- (18.25,7.75)node[pos=0.5,above, fill=none]{p};
\draw [ color={rgb,255:red,0; green,30; blue,255}, ->, >=Stealth] (5.75,7.75) -- (8.75,7.25);
\draw [ color={rgb,255:red,0; green,30; blue,255}, ->, >=Stealth] (8.75,7.25) -- (11.5,7.25);
\draw [ color={rgb,255:red,0; green,30; blue,255}, ->, >=Stealth] (11.5,7.25) -- (14.25,7.25)node[pos=0.5,below, fill=none]{$p^{Ø}$};
\draw [ color={rgb,255:red,0; green,30; blue,255}, ->, >=Stealth] (14.25,7.25) -- (16.25,7.5);
\draw [ color={rgb,255:red,0; green,30; blue,255}, ->, >=Stealth] (16.25,7.5) -- (18.25,7.75);
\draw [line width=1.5pt] (10,10.25) -- (10,7.75)node[pos=0.5,left, fill=none]{$h_\perp$};
\draw [line width=1.5pt] (10,7.75) -- (13,7.75)node[pos=0.5,above, fill=none]{$h_\parallel$};
\draw [dotted] (13,10.5) -- (13,7.75)node[pos=0.5,right, fill=none]{$d$};
\node [font=\large] at (5.5,7.75) {O};
\node [font=\large] at (18.5,7.75) {D};
\end{circuitikz}
}%
\caption{Between O and D we construct: a straight line $s$ (dashed red), the FP $p^{Ø}$ on the empty graph (blue), and the FP $p$ for the loaded graph. Since $p$ contains a dysfunctional edge, $p^*$ is the longest allowed path.}
\label{fig:path-features}
\end{figure}

We characterize the morphology of a path from O to D (with $s=|OD|$ being the straight-line distance) by looking at two geometrical properties:
\begin{itemize}
    \item Detour $d = \max_{i\in p}h^i_{\perp}$: maximum distance between the path and the straight line;
    \item Inness $I \approx \sum_{i\in p}h^i_{\perp}h^i_{\|}$: signed area between the path and the straight line (computed using the trapezoidal rule);
\end{itemize}
where $h^i_{\|}$ stands for the scalar product of the $i$-th directed edge with $\overrightarrow {OD}$, while $h^i_{\perp}$ is the edge distance from $\overline{OD}$. Fig.~\ref{fig:path-features} offers a visual definition of these objects.

The sign of the computed area is taken to be positive for paths traveling toward the center of the map, negative otherwise: Thus, it will be positive whenever a ``centripetal force''~\cite{lee_morphology_2017} exists. Normalized $I$ is obtained by division with the square built on the straight line: $\tilde I=I/s^2$ and normalized detour as $\tilde d=d/s$.

\subsection{Slowdown, Completeness and Performance Index}
Given the network in some generic state and an OD pair, the fastest path $p$ (composed of a sequence of edges of length $l_i$) is computed. A special case of $p$ is the FP as computed over the network in its empty state: $p^{Ø}$.
We define the Completeness factor ($C$) as the ratio between the distance $|p^*|$ that can actually be traveled along $p$, and the whole path length $|p|$:
$$C^{OD}=\frac{|p^*|}{|p|}=\frac{\sum_{i\in p^*}l_i}{\sum_{i\in p}l_i}.$$
The admissible traveling distance over the path is either limited by the presence of unavoidable dysfunctional edges (i.e., the network is fragmented) or because $\tau$ would be exceeded. In both cases, we load the network only for the reachable part of the path $p^*$.

The Slowdown factor ($S$) is ideally defined as the ratio between the travel time from O to D on the FP over the empty network $T_{p^{Ø}}=\sum_{i\in p^{Ø}}t_i^{Ø}$, and the one necessary on the FP available (typically a different one) on a congested system $T_p=\sum_{i\in p}t_i$. A unity $S$ means full speed, whereas $S\rightarrow0$ is found in congested edges. This simple definition must be slightly modified to take into account those paths that cannot be completed either because of disconnected O and D (hard reject condition) or because travel time would be longer than the chosen $\tau$ (soft reject condition). In these situations, to be meaningful, the slowdown computation is limited to the portion of the path that is possible to travel over the congested network by rescaling the time on the empty network by $C^{OD}$:
$$S^{OD}=\frac{T^{OD}_{p^{Ø}}\cdot C^{OD}}{T^{OD}_{p^*}}.$$
$T_{p^{Ø}}$ is the total travel time along the FP $p^{Ø}$ between $O$ and $D$ when the graph is empty. $T_{p^*}$ is the travel time of the OD path until the dysfunctional edge in the case of the loaded graph. $T_p>\tau$ values are clipped.  

Finally, we introduce the \emph{Performance Index} ($P$) for a path connecting O and D as the product of $S^{OD}$ and $C^{OD}$ for any network state:
\begin{equation}
    \label{eq:perf_index}
        P^{OD}=S^{OD} \cdot C^{OD}.
\end{equation}

In a low congestion regime, $P$ will simply follow the slowdown behavior, whereas for high traffic volumes, the completion ratio will often dominate, with several paths containing dysfunctional edges preventing the transport from reaching its destination. The $P$ index is a non-local performance metric that evaluates the efficiency of the transport over the paths as opposed to standard metrics that focus on the performance of the single network constituents such as edges and nodes. Transport performance is heavily influenced by the kind of process taking place over the network and induces a specific spatial correlation on edge usage that no uniform averaging over the edges can approximate. Our choice of the $C$ factor is the only part of this definition that is tailored for vehicular traffic since it assumes that a partially completed transport is still valuable.
%

\subsection{Non-uniform path performance degradation}
The Gini coefficient has been used in recent years to describe structural properties of networks, such as the inequality of the degree and BC distributions~\cite{ding2018detecting}, and to extract structural profiles of urban networks~\cite{diet_towards_2018}.
We compute the Gini coefficient of the $P$ distribution to evaluate inequalities in path performance degradation. To measure inequality in a set of quantities $\{P_i\}$ we define:
$$G=\frac{\sum_i^n\sum_j^n |P_i-P_j|}{2n\sum_i^n P_i}',$$
which tends to zero (perfect equality) when most paths have similar $P$ values; on the other side of the spectrum, maximum inequality among $P$ values is signaled by $G\rightarrow1$: i.e., only a few paths maintain acceptable performance while the rest is dysfunctional. 

\section{Results and discussions}
\label{sec:results}
We selected eight large metropolitan areas: Beijing, Berlin, Las Vegas, London, Los Angeles, Madrid, Paris, and Rome to represent very different topologies, mainly stemming from different geographical locations and historical evolution processes. The vehicular transport network of the areas around city centers ($20$ km radius except for Rome and Madrid, with $12$ and $15$ km, respectively) were downloaded from OpenStreetMap (OSM) retaining the information about edge length, maximum speed, and number of lanes (see tab.~S1 for details). The number of edges of the corresponding graphs is in the range $1.0\times10^5$ (Rome) to $5.5\times10^5$ (London). The maximum traffic load we generated for all cities was limited to $2.0\times10^6$ OD pairs, enough to drive all cities into a deeply congested state. The maximum allowed time for a path to be completed was set to $\tau=3600$ s. All vehicles travel at the maximum speed allowed on the edges, depending on the state of the simulation at a specific time. In contrast to the simulations on the same graphs performed for a previous publication~\cite{cogoni2024predicting}, we do not perturb speed values with Gaussian noise to simplify the analysis\footnote{We experimentally verified that results are not affected in relevant ways.}. To estimate the variability of the phenomenon, we run, for every city, $10$ simulations, each with an independently sampled OD set.

\subsection{Path shape analysis}
We perform an analysis of the shape of the paths modeled after previous works on spatial networks~\cite{kartun2019shape,lee_morphology_2017}.
Each path is characterized by the following geometrical properties and their scaling with the straight-line OD distance:
\begin{itemize}
    \item path travel time $T$ (effective length);
    \item \emph{detour} ($d$): maximum distance along the path from the straight line between O and D;
    \item \emph{inness} ($I$): the area comprised between the path and the straight line, considered as positive when leaning toward the center of the map.
\end{itemize}
We study how the above properties change as the network is subject to increasing traffic levels. The scaling behavior with the OD distance is studied in a log-log space by performing a linear fit to extract the ``critical'' exponents. A strict definition of these exponents holds only when considering infinite networks, but the idea is useful to characterize real finite networks~\cite{li_percolation_2015}. We focus on the scaling exponents for the standard deviation of travel time ($\sigma(T)\sim s^\chi$), and for the wandering ($\mathbb E(d) \sim s^\xi$). We also compute the exponent $m$ associated with $T$ scaling ($\mathbb E(T) \sim s^m$). We denote the average of a quantity by $\mathbb{E}(\cdot)$ and $a\sim b$ means $a$ converges to $Cb$ with $C$ a constant independent of OD, as $s\rightarrow\infty$.
\subsubsection{Critical exponents in the low-congestion regime}

\begin{figure}[htb]
    \includegraphics[width=\linewidth]{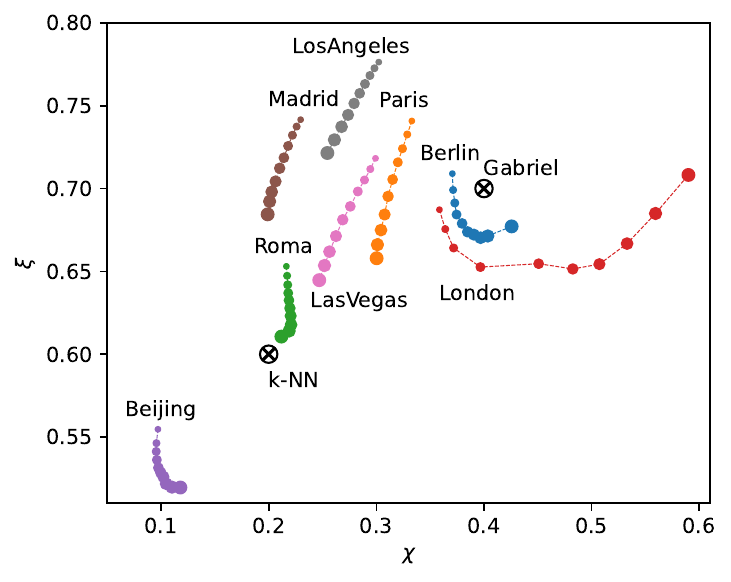}
    \caption{\label{fig:chi_xi_clusters}Critical exponents $\chi$ and $\xi$ for all cities (and for the Gabriel and k-NN graphs) for traffic levels up to $250$k vehicles.}
\end{figure}

Path travel time scaling ($m$) is well approximated by a linear fit in the log-log space for all cities except for Beijing where the curve does not follow a power law. The linear fit holds up only up to $\sim3$ km for low traffic, while linearity is regained deep into the congested regime. The results of the linear fits performed with increasing traffic (below the critical level) for real cities and for two random planar graphs (Gabriel and k-Nearest Neighbors (K-NN)) are reported in~\cref{fig:chi_xi_clusters}.

\begin{table}[!ht]
        \begin{tabular}[t]{l@{\hskip .1in}r@{\hskip 0.1in}r@{\hskip 0.1in}r}
            \hline
            city & $\chi$ & $\xi$ & $m$\\
            \hline
            \noalign{\vskip 1mm}
            Beijing & $0.10$ & $0.55$ & $0.73$ \\
            Berlin & $0.37$ & $0.71$ & $0.76$ \\
            Las Vegas & $0.30$ & $0.72$ & $0.75$ \\
            London & $0.36$ & $0.69$ & $0.76$ \\
            Los Angeles & $0.30$ & $0.78$ & $0.78$ \\
            Madrid & $0.23$ & $0.74$ & $0.72$ \\
            Paris & $0.33$ & $0.74$ & $0.74$ \\
            Roma & $0.22$ & $0.65$ & $0.76$ \\
            Gabriel & $0.40$ & $0.70$ & $1.00$\\
            k-NN & $0.20$ & $0.60$ & $1.00$\\
            \hline
        \end{tabular}
\caption{\label{tab:chi_xi_m}Critical exponents $\chi$, $\xi$, and $m$ computed for all cities with no traffic and for the undirected Gabriel (and k-NN) graph ($\rho=10^{-4}$). Linear fits, and their complete evolution with growing traffic, are shown in figs.S17-S24}
\end{table}

The values of $\chi$ and $\xi$ shown in~\cref{tab:chi_xi_m} are drawn for all cities in~\cref{fig:chi_xi_clusters} as the smallest circles of each city and represent the empty network state that increases in size up to $250$k vehicles.
The Gabriel (defined by excluded region) and the k-NN (defined by node proximity) graphs are also included as a reference in their undirected versions with densities comparable to real cities (mean edge length about $80$ m). London shows a significant variation of the fitted exponents already with the addition of a few thousand vehicles (because its transition to congestion happens earlier at $\sim0.3$M vehicles) while other cities are more stable in this low traffic regime with about half the drift: This drift involves in all cases a decrease of both $\xi$ and $\chi$ for all cities but London and, to a lesser extent, Berlin (increase of $\chi$ and later $\xi$ inversion), and Rome (almost constant). London and Berlin are very close to Gabriel, and Rome ends very near the k-NN. Beijing has the lowest values for both exponents ($\chi\sim0.1$ and $\xi\sim0.55$) while Los Angeles, Las Vegas, Paris, and Madrid have intermediate $\chi\sim0.3$ and the highest $\xi\sim0.75$ initial values. A coherent comparison of real cities with synthetic graphs involves directional edges for the Gabriel and k-NN and some form of sparsification (to remove perfect bidirectionality) and will be the subject of a future work. Critical exponents within the congested phase are not discussed since the fits are noisy (especially $\chi$) with large error bars beyond the transition: their full behavior is visible in figs.~S17-S24, though.

\subsubsection{Morphological changes and central force effects}

\begin{figure}[htb]
    \includegraphics[width=\linewidth]{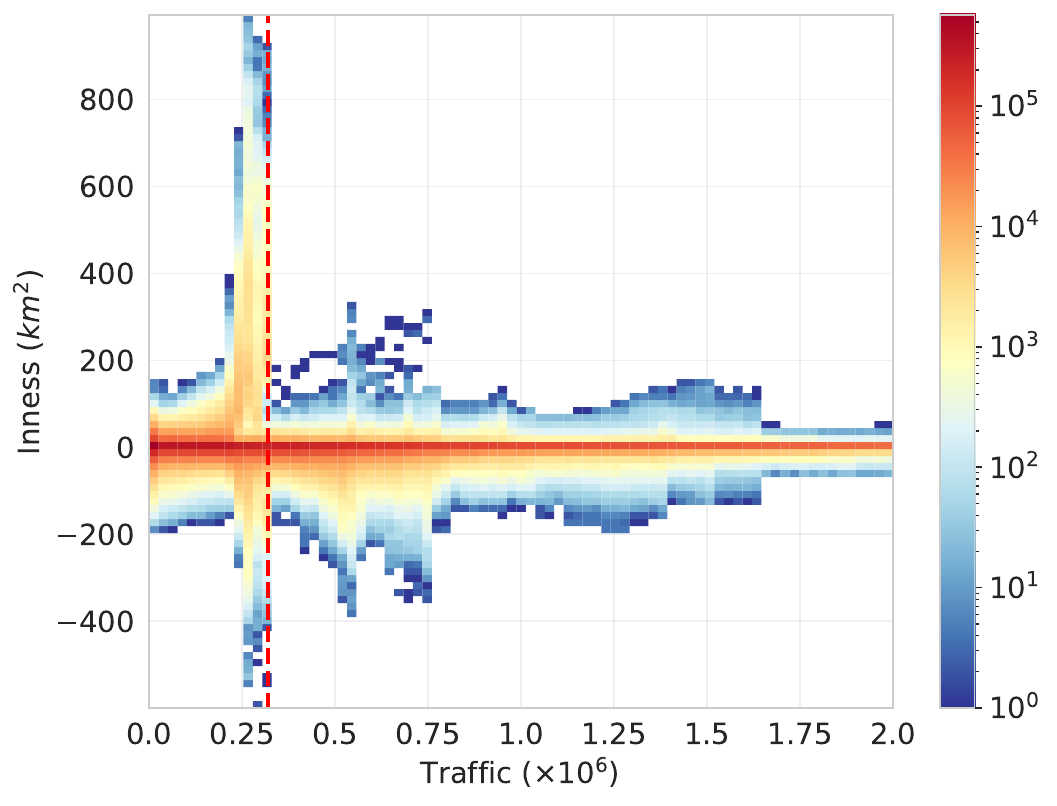}
    \includegraphics[width=\linewidth]{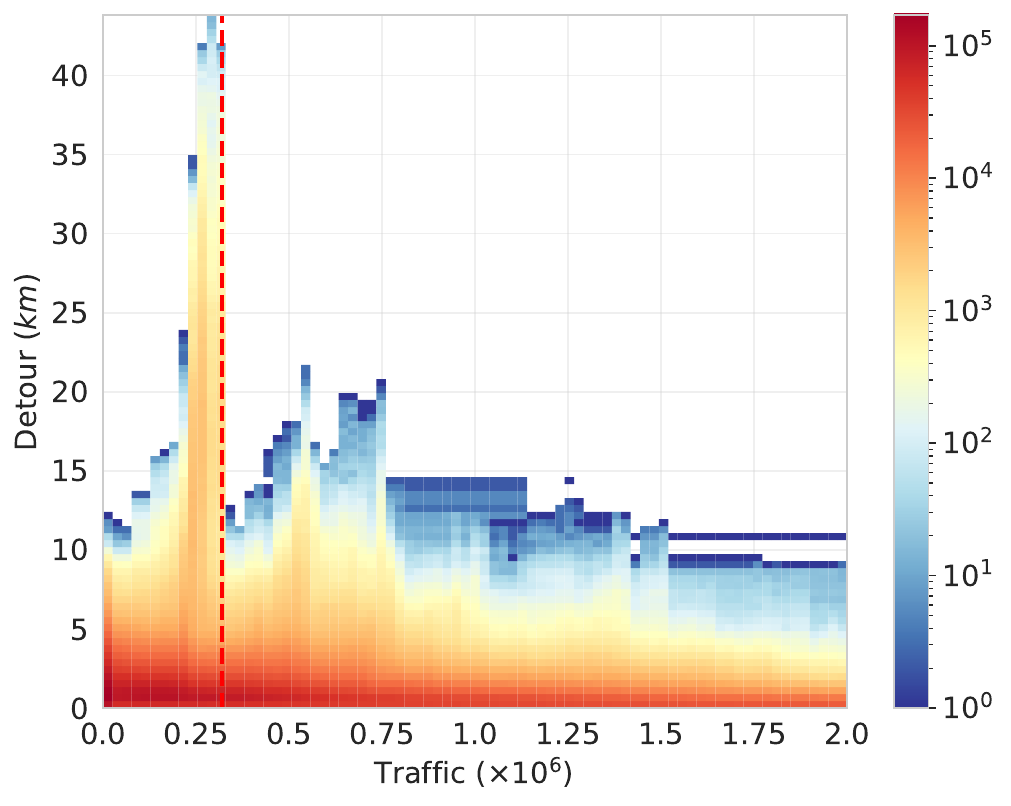}
    \caption{\label{fig:inness_detour_distr}The full distributions of inness and detour for London. The logarithmic color bar represents how many ($\#$) paths fall in each bin, with blue for $\#<10$ and red for $\#>10^2$. The congestion transition is shown by vertical dashed red lines. Each column contains $\sim5\times10^5$ vehicles.}
\end{figure}

\begin{figure}[htb]
    \includegraphics[width=\linewidth]{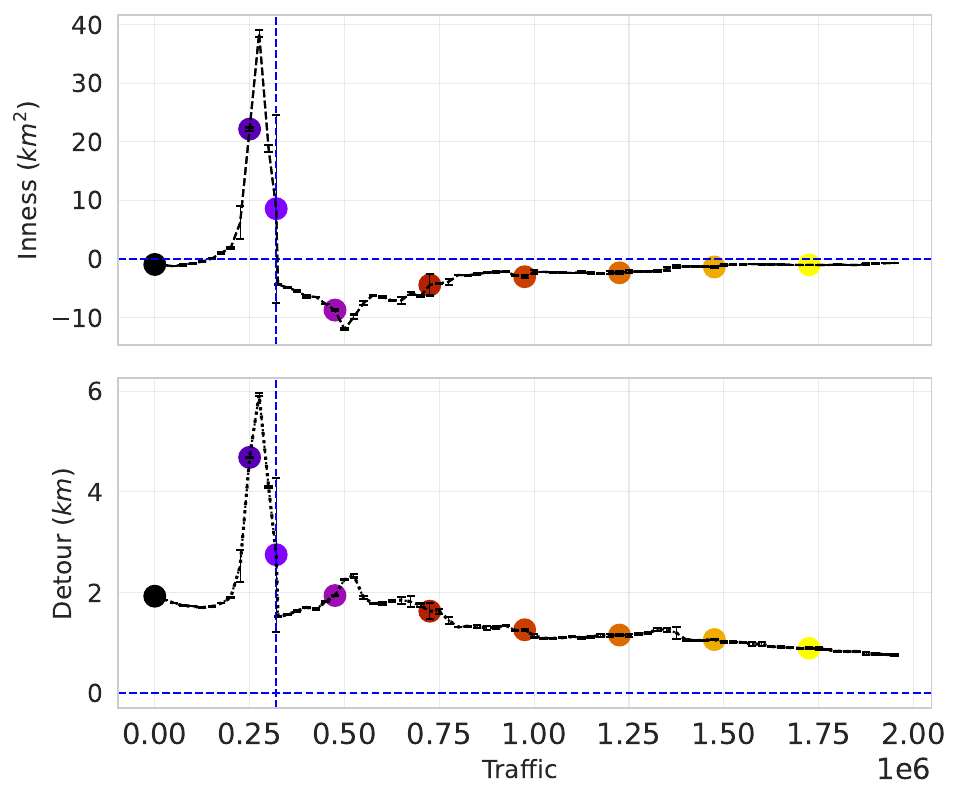}
    \caption{\label{fig:inness_detour_mean}Mean values of inness and detour for London. Small colored circles are reference traffic volumes used later. The congestion transition is shown by vertical dashed blue lines. Colored circles following the \emph{``gnuplot''} colormap serve as a guide for traffic levels later used in~\cref{fig:london_wandering1}.}
\end{figure}

The $\xi$ and $\chi$ exponents describe the scaling of the detour and path-length variability on average, but more information may be extracted from their full distributions at all traffic levels. For the city of London, the two distributions
$\mathcal{P}(I)$ and $\mathcal{P}(d)$ are shown in~\cref{fig:inness_detour_distr} where the color indicates the number of paths in each bin. The inness and detour distributions are wide and non-Gaussian in general, but we also report their average values in~\cref{fig:inness_detour_mean} to better visualize where the majority of paths is located.

\begin{figure}[htb]
    \includegraphics[width=\linewidth]{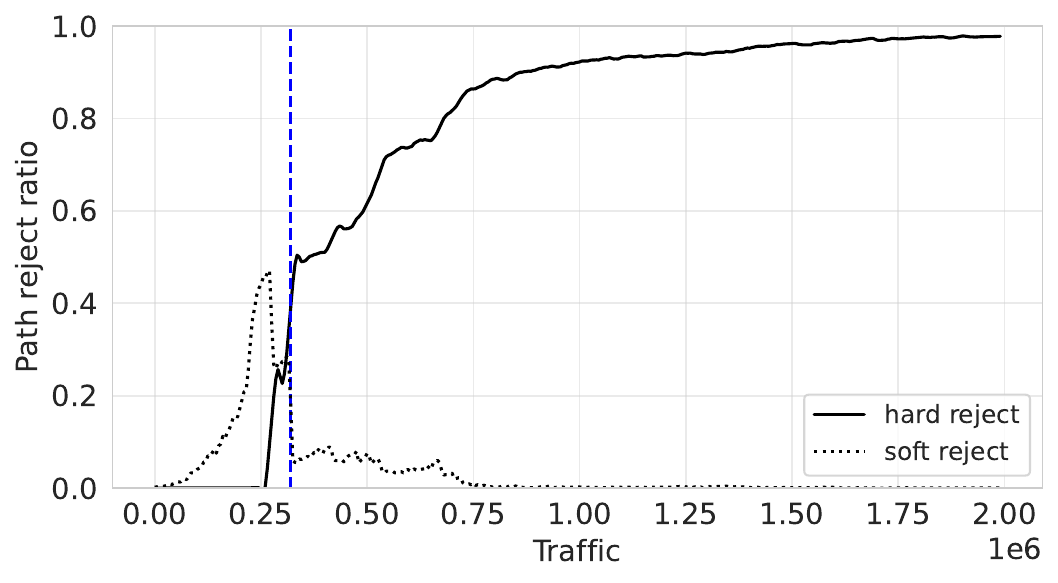}
    \caption{\label{fig:path_reject_ratio}Path reject ratio: solid black line are the hard-rejects (no path without dysfunctional edges available), dotted line are the soft-rejects whose travel time exceeds $\tau$ and can only complete a fraction of their transport.}
\end{figure}

For low traffic levels (below $0.2\times10^6$ vehicles) the detour distribution is limited to a maximum of about $10$ km over the whole network, with most values below $5$ km. Increasing traffic toward the transition (red vertical dashed line at about $0.3\times10^6$ vehicles) the distribution suddenly expands to $d>40$ km for a relevant number of paths that are trying to avoid preferential (faster, shorter, multilane) roads that have become congested. This transition traffic volume (clearly specific for each network) coincides also with the position of the flex in the curve of the hard-rejected path ratio in~\cref{fig:path_reject_ratio} (paths in hard reject condition). A peak in the soft reject ratio (paths in soft reject condition) seems to be a reliable precursor of this abrupt change in the network transportation performance, immediately followed by the hard reject sudden growth. For London, the soft reject ratio reaches about $50\%$ of all paths being added with a traffic volume of about $0.25\times10^6$ vehicles. It should be noted that the soft rejection depends on the relation between the size of a city and the maximum travel time $\tau$, and it appears that the standard choice of $\tau=3600$ s may be too long for the smaller urban areas (e.g., Rome and Madrid) that produce soft-rejection peak ratios under $10\%$.
Each steep growth in the hard rejection ratio signals a break-up of the network into multiple disconnected islands. In this case, it is interesting to note that having $50\%$ of hard-rejects after the main transition (see \cref{fig:path_reject_ratio}) roughly tells that the bridges over the Thames have collapsed as will be confirmed in the following.
The fraction of edges needed to produce a breakup in the connectivity is typically very small ( $\lesssim0.1\%$, see figs.~S1-S4(bottom)) since the first edges to reach maximum saturation are the most desirable ones, on average.

While detour $d$ highlights the maximum distance of the path from the straight line, another way to quantify a shape change in a path is to compute its inness. Since the sign is positive-definite for paths leaning toward the map center, we can measure the effects of the central force due to topology or congestion of the city center~\cite{lee_morphology_2017}. In~\cref{fig:inness_detour_mean}(top) we see how the average inness $I$ evolves with traffic intensity: For an empty network, the value is very small and slightly negative (centrifugal). Increasing traffic, the central force disappears to later become centripetal, then exploding in a huge peak just before the transition for which $I$ reaches a mean value of $40$ km$^2$, but some paths are jolted to the other side of the network (across the center, thus the inness positivity, but an effectively repulsive behavior nonetheless) reaching maximal values of more than $I\sim800$ km$^2$. The transition coincides with the maximum path-choice entropy (flatter cost-of-transport surface~\cite{akamatsu1997_path-choice_entropy,cogoni2017ultrametricity}) with paths spanning the entire network in their selfish struggle to finish within $\tau$ (or finishing at all). The peaking of $d$ and $I$ happens exactly at the same traffic level, just before the transition.
After this short-lived transitional regime (involving the addition of about $0.1$M vehicles), we observe an abrupt decrease of $I$ and a crossover to moderately negative values: in this regime, the network center becomes repulsive and stays this way up to very high congestion. Continuing to increase traffic, it is evident that the magnitude of the areas shrinks with $I\rightarrow 0$. This is due essentially to a large portion of proposed paths being hard-rejected as seen in~\cref{fig:path_reject_ratio}. The slowest paths are rejected first, then only the fastest ones can be added to the network without O and D belonging to different network islands. The same fate happens to detour with $d\rightarrow0$ for the same reasons. Detour and inness distributions for all cities are visible in figs.~S5-S8 with similar results, but generally showing a slightly attractive central force for all cities except for Rome in which the center strongly repels all paths but those with small $s$. Los Angeles and Las Vegas have, on the other hand, fairly attractive centers when empty.

\begin{figure*}[htb]
    \includegraphics[width=\linewidth]{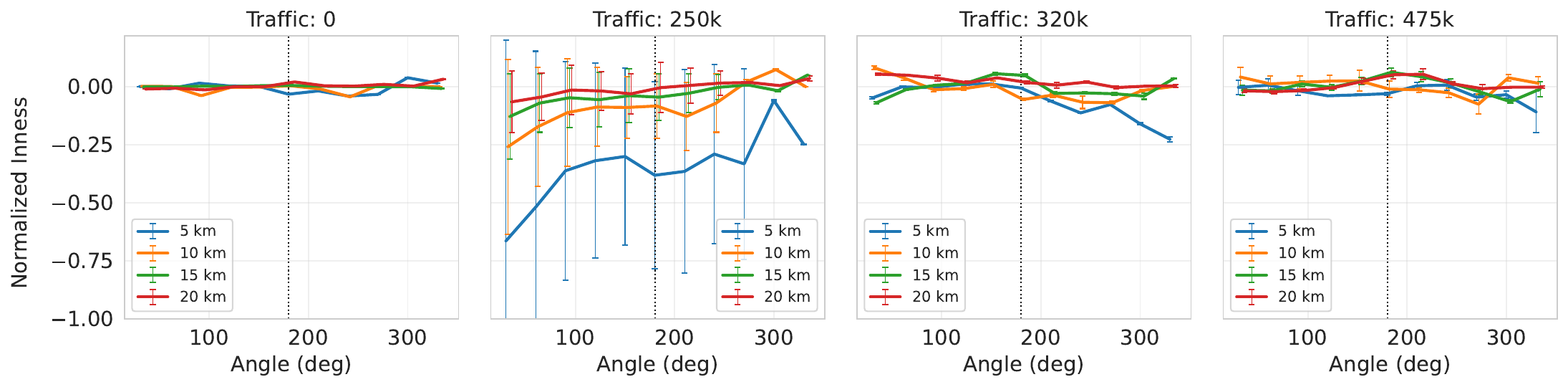}
    \caption{\label{fig:norm_inness}Normalized inness for four traffic levels (the third plot corresponds to the transition traffic level). Vertical dotted line divides CW from CCW paths: $330^{\circ}$ means $30^{\circ}$ CCW}
\end{figure*}

A normalized inness $\tilde I=I/s^2$ (divided by the squared straight-line distance) allows us to directly compare the effects of the central force on paths of different lengths at several traffic levels. We study $\tilde I$ for several angular separations and distances of ODs from the center of the map to gather a better understanding of path shape dependence for specific OD configurations and traffic levels. In~\cref{fig:norm_inness} we plot $\tilde I$ at four traffic levels for our reference city of London: Empty, $250$k, $320$k (the flex position), and $475$k vehicles. With an empty network (left plot), the average $\tilde I$ is very small for all path configurations, thus no central force effects are detected at any distance from the center in this case. Approaching the transition traffic level (second plot), the situation suddenly changes with the most dramatic effects seen at small angular ($\alpha$) separations ($5$ km, blue line): for $\alpha\in(30^{\circ},90^{\circ})$ the repulsive effect is strongest, with the average path being pushed away from the center to form an area of about half the OD square with the straight line $\tilde I\sim-0.5$. This effect shows a chiral preference since for $\alpha\in(270^{\circ},330^{\circ})$ (equivalent to the previous separations, but counterclockwise) the effect is much smaller with $\tilde I\sim-0.2$, but still repulsive. Error bars are very wide for all $\alpha$ except for $300^{\circ}$ and $330^{\circ}$ ($30^{\circ}$ and $60^{\circ}$ CCW). For a larger radius (orange and green) the effect is weaker, but still repulsive for all angles and the error bars get smaller. At $20$ km (red), $\tilde I\sim0.0$ except for very small CW angles. Beyond the transition (i.e., $475$k) distributions become very narrow with error bars within symbol size and a small repulsive effect remains only for $5$ km radius and small angle CCW paths. The results regarding the empty networks, the only ones we can directly compare with our city selection, are quite different than the average ones published by Lee et al.~\cite{lee_morphology_2017}: 
Figs~S9-S16 contain the same plots for the other cities.

\begin{figure*}[htb]
    \includegraphics[width=.49\linewidth, trim={1.5cm 0.7cm 1.5cm 1cm}, clip]{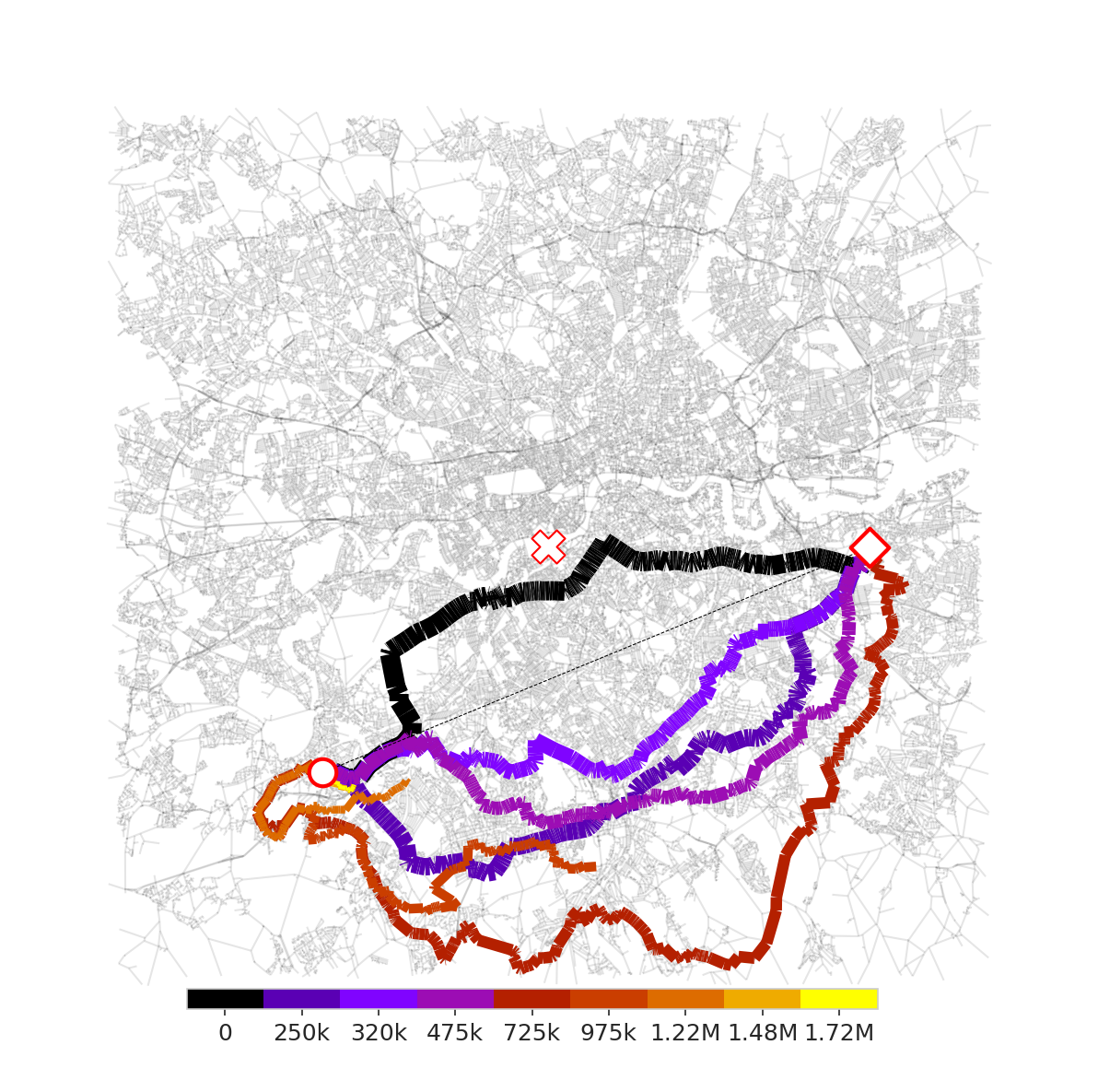}
    \includegraphics[width=.49\linewidth, trim={1.5cm 0.7cm 1.5cm 1cm}, clip]{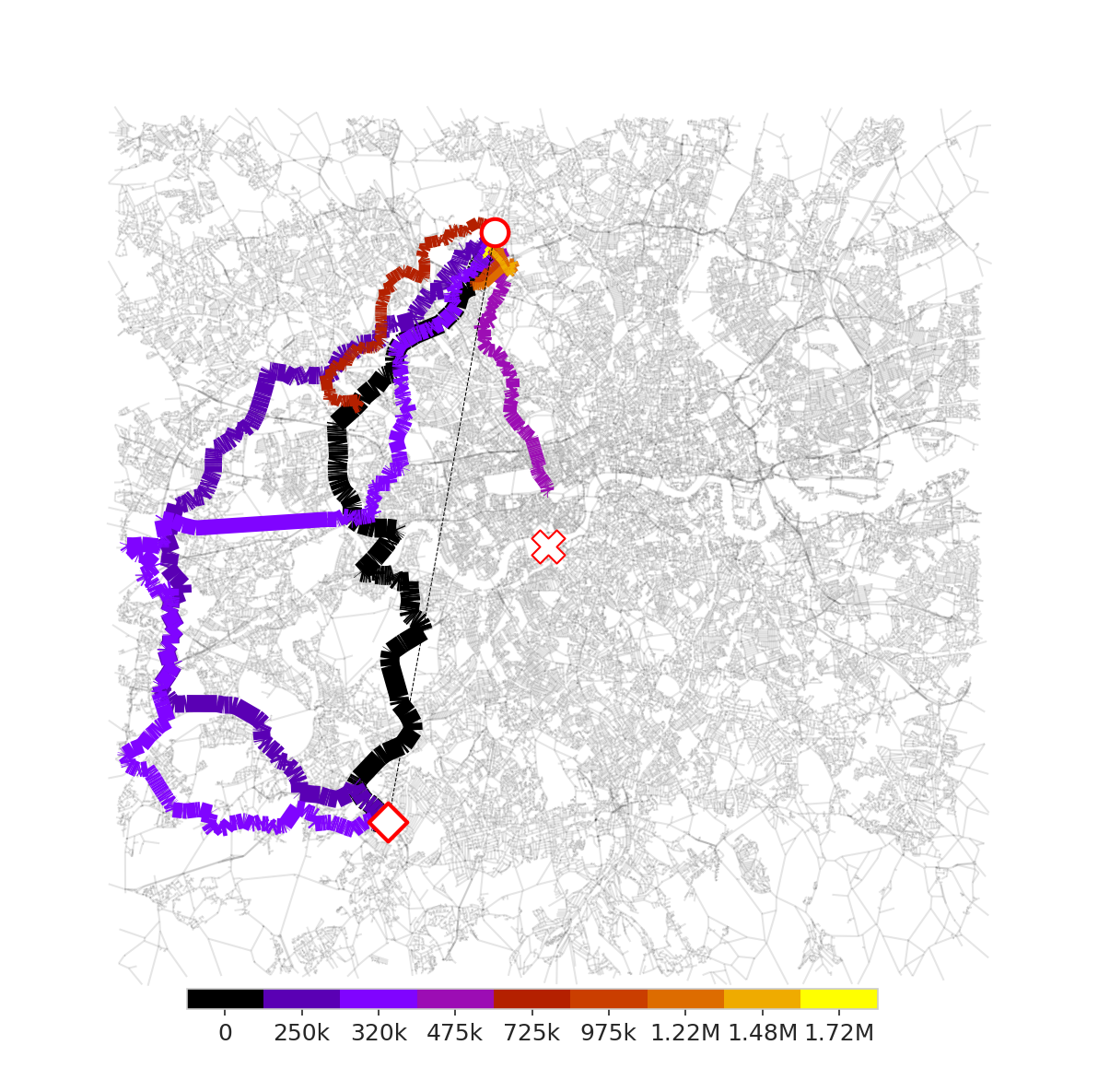}
    \caption{\label{fig:london_wandering1}In both panels: FPs between same OD (both at $15$ km radius) at different traffic load. (The circle and diamond symbols refer to Origin and Destination respectively, whereas the cross stands for the center of the map). On the left O and D are both below the river, while on the right the river must be crossed. The color bar shows the traffic level for each path. More angular configurations for all cities are visible in figs.~S25-S32.}
\end{figure*}

To get a better idea of what happens to the FPs while increasing traffic, we show two examples in~\cref{fig:london_wandering1} for the Greater London area that share similar OD separations but differ in orientation: In the left panel, the vehicle starts from south-west (red circle) to reach the east side (red diamond). Both O and D are south of the River Thames and the wide black path is the FP in the empty state: in this case, it is attracted toward the center with respect to the straight line (dotted gray line). As traffic grows (violet to red) the force becomes strongly repulsive, peaking at about thrice the transition traffic when the red path is pushed toward the network's southeast corner. Beyond this traffic level, no path exists for this OD and only a very small part of the journey (orange and yellow) can be traveled: this is expected from \cref{fig:path_reject_ratio} since about $85\%$ of paths are rejected.
On the right panel, we consider a north-south orientation with O and D on opposite sides of the river: the empty-network FP almost follows the straight line (black), but gets repelled from the center at the transition traffic (violet). Beyond the transition, no available full paths exist (even the purple path stops very early). This breakup happens with three times less traffic than before because the river acts as the weakest part of the urban network connectivity, as expected.

\subsection{Slowdown, Completeness and Performance Index}
\begin{figure}[htb]
    \includegraphics[width=\linewidth]{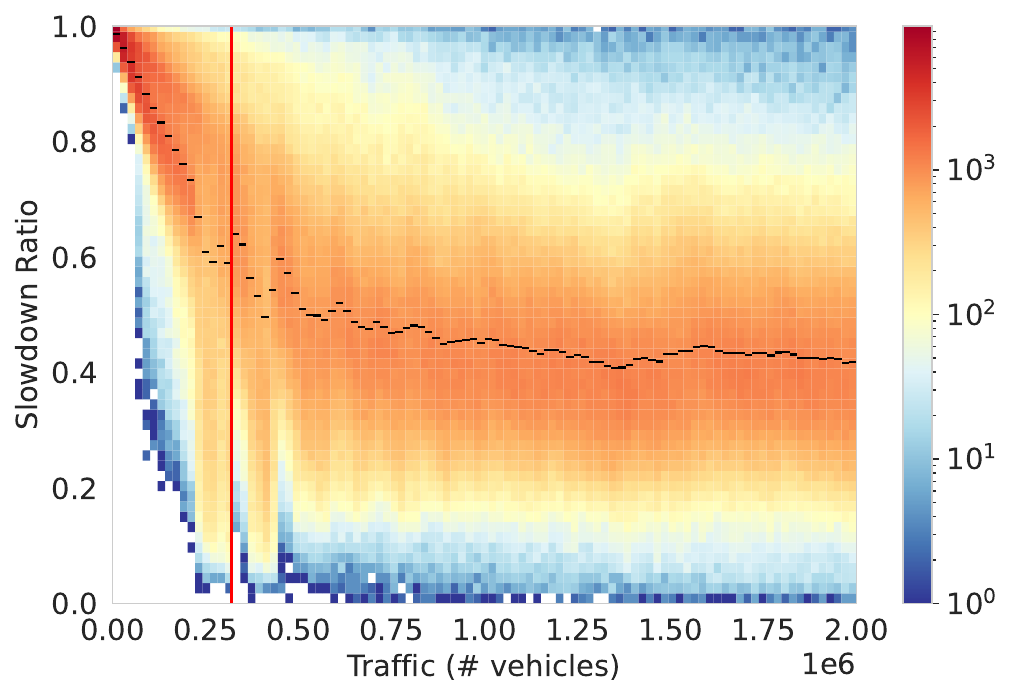}
    \includegraphics[width=\linewidth]{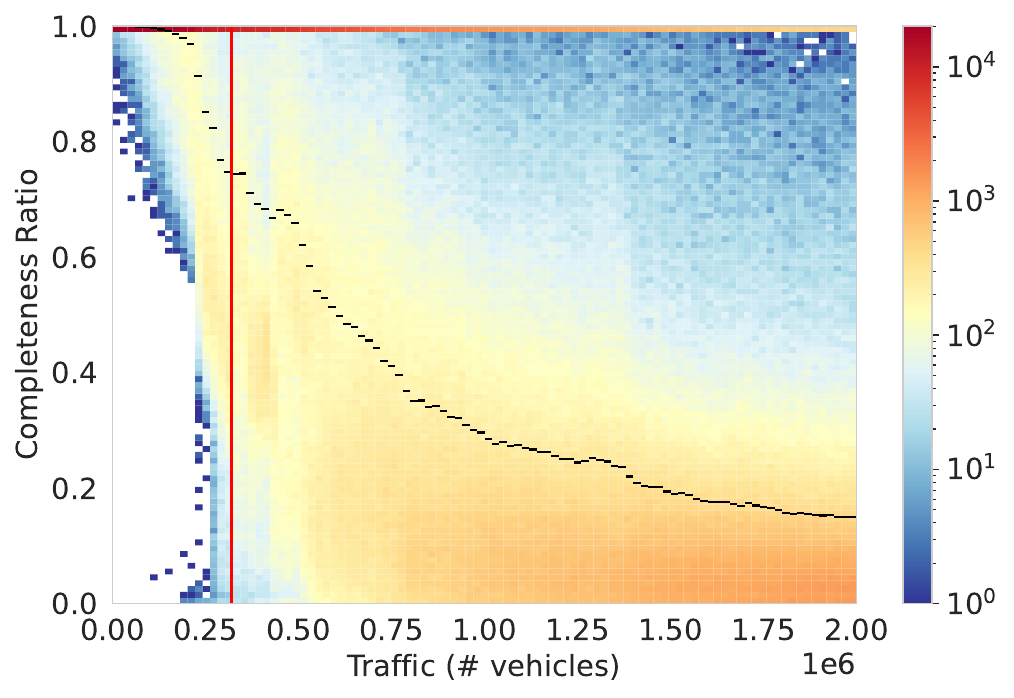}
    \includegraphics[width=\linewidth]{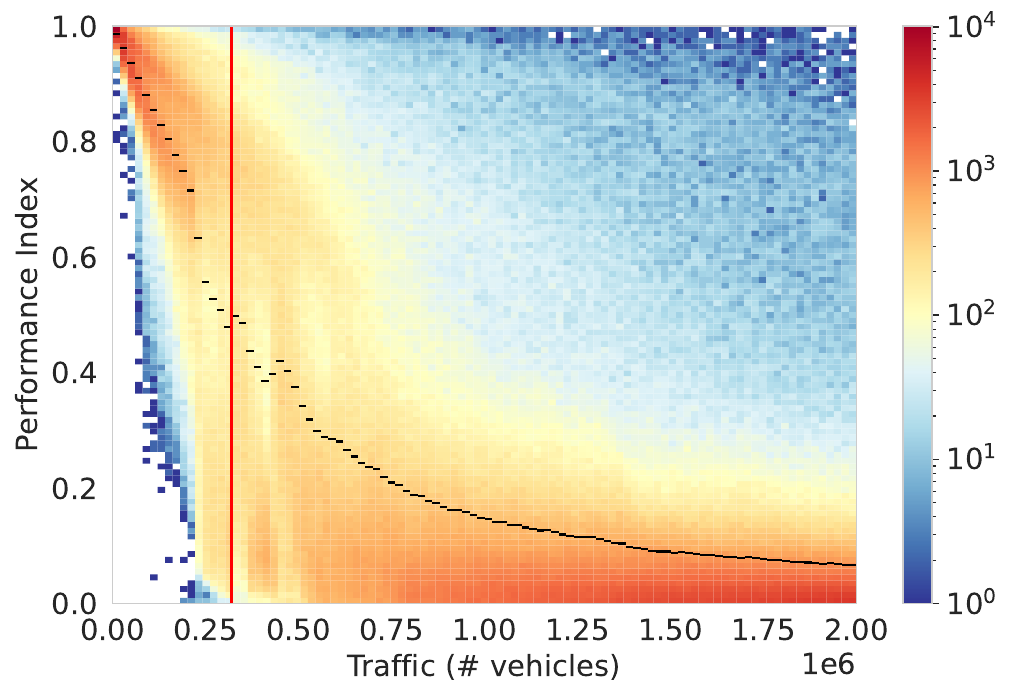}    
    \caption{\label{fig:london_s_c_p}Distributions (counts of paths per bin) of $S$, $C$ and $P$ for London for increasing traffic. The red line is the transition and the black curve is the average for each traffic bin (containing $25$k vehicles). See figs.~S33-S36 for the other cities.}
\end{figure}

In previous sections, we saw how the shape of FPs depends on the congestion state of the underlying network and on the OD positions with respect to the map center.
We now turn our attention to how shape impacts path performance at increasing traffic levels. Our simple traffic model linearly connects edge density to the slowdown experienced by vehicles on average. Thus, the most obvious effect that happens to vehicles when increasing the traffic level starting from an empty network is a degradation of travel times. This slowdown appears first on the most desirable roads (high speed, multi-lane, few traffic lights) since everybody chooses them as a first choice, especially for longer distances. As congestion starts to cripple the performance of these main roads, the FPs available necessarily switch to lesser-used (and capable) infrastructures. In~\cref{fig:london_s_c_p}(top) we see how the population of paths connecting uniformly distributed OD pairs all over the network experiences slower speeds as the traffic volume increases. With zero traffic all paths are concentrated in the top left corner ($S\sim1$) meaning maximum speed; as traffic grows, some paths slowdown and the distribution smears over wider values, with some OD at half their empty speed already with $100$k vehicles, but most of them (average $S$ of vertical bins shown by horizontal black bars) maintain values $>0.8$. Right before the transition (red vertical line), the slowdown distribution widens spanning the whole $[0,1]$ range with an average value of $\sim0.6$, with a non-negligible fraction of paths having one-tenth of their empty speed. A complex structure of oscillations for the $S$ variance is evident from $0.2$M to $0.5$M happening in phase with further network breakups also visible in \cref{fig:path_reject_ratio}: we can say that each ``sub-transition'' (with its smaller sigmoid/flex pattern) induces a new broadening of $S$ and smaller peaks for detour and inness values (\cref{fig:inness_detour_mean}).
Slowdown then stabilizes at $\sim0.4$ with a wide distribution until the maximum load ($2$M vehicles): nothing dramatic happens in this regime since the addition of longer paths is prevented, so almost only very short and partial paths are added.
This fact brings us to the next quantity of interest to characterize travel efficiency: Completeness, which measures, for each OD pair, how far it is possible to travel along the FP. In urban networks, it makes sense to take into account this factor because between being able to travel $90\%$ or $50\%$ of the path results in a very different outcome from a traveler's perspective. In other contexts, a non-linear behavior could be better suited, with the extreme case of all or nothing, such as in digital communication networks, where a data packet reaching $99\%$ of its path is indistinguishable from one stopping near its origin. Even brewing espresso coffee would not gain much from water paths not fully reaching the lower side of the filter.

In this urban network context, for the city of London, we see that $\overline{C}$ (black horizontal lines) drops from $1.0$ steeply after a short plateau, and just before the transition, already one-third of the paths cannot reach its destination, either because $T_p>\tau$ or because only dysfunctional paths exist. The thin stripe of red bins at the top of~\cref{fig:london_s_c_p}(middle), indicating completed paths, progressively becomes fainter and most paths prematurely stop after $1$M.

In this context of urban transport makes sense to multiply $S$ and $C$ to obtain an overall estimate of how fast, and how far each path was able to go on its OD journey. We call this Performance Index $P$ and we can see that its distribution is modulated by the $S$ and $C$ factors above it in~\cref{fig:london_s_c_p}: The ${\overline P}$ decrease is almost exponential with traffic and shows some non-trivial structures near the transition due to the succession of breakups of different parts of the city network. Beyond the transition, paths worsen their $P$ from $0.4$ to less than $0.1$. A similar trend emerges for the other cities: see figs~S33-S36.
This behavior demonstrates that once we reach the vicinity of the transition, the overall performance of a realistic transport phenomenon deteriorates very quickly with small traffic variations: we can say that the network has a large susceptibility in this regime that we could define as the derivative of $P$ with respect to traffic volume: ${\partial P}\over{\partial V}$. 

\begin{figure*}[htb]
    \includegraphics[width=\linewidth]{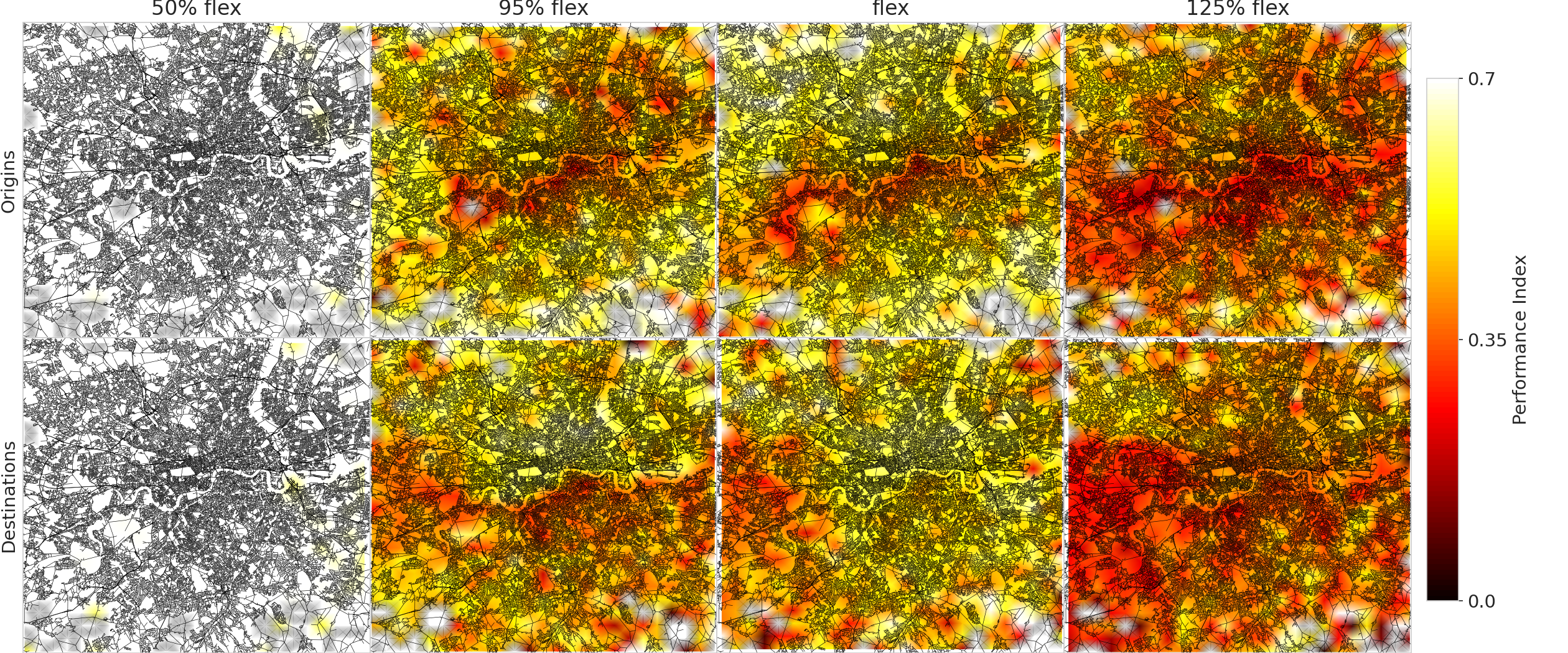}
    \caption{\label{fig:london_p_map}$P$ maps for London: Top (bottom) plots show the average $P$ value associated with the Origin (Destination) nodes of each path.}
\end{figure*}

\begin{figure}[htb]
    \includegraphics[width=\linewidth]{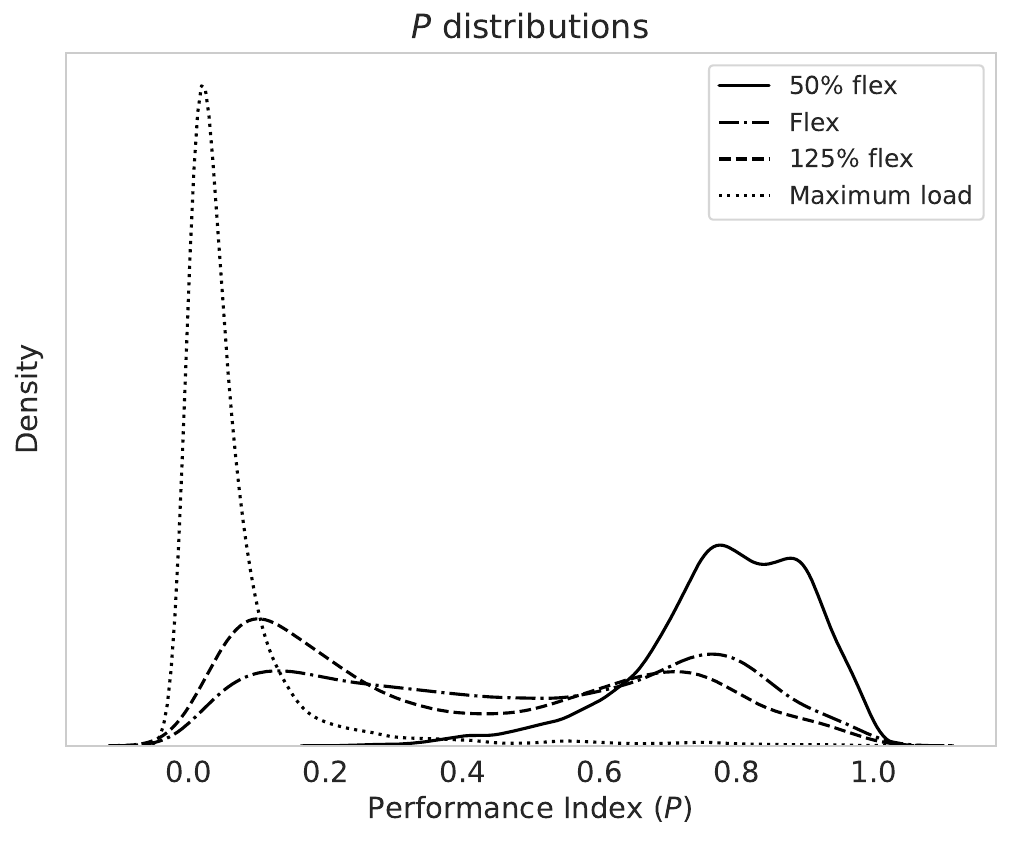}
    \caption{\label{fig:london_pi_distr}London $P$ distributions at four traffic levels, three of which can be directly compared to the $P$ maps in~\cref{fig:london_p_map}.}
\end{figure}

It is very useful to take the same $P$ distribution for all paths at specific traffic levels (four vertical slices of~\cref{fig:london_s_c_p}(bottom)) and map each path's performance onto its origin~\cref{fig:london_p_map}(top row) and to its destination~\cref{fig:london_p_map}(bottom row) on the topographic chart. In this case, the color of a map location represents the average $P$ of all paths leaving from (and going everywhere else) or arriving there from all other places.
The left column shows the lowest traffic ($50\%$ of the transition flex) when acting as traffic sources (O, top) and destinations (D, bottom). The maps are almost unmarked in this case since most paths perform very well with $P>0.7$. Moving toward higher traffic ($95\%$ of flex), $P$ decreases fast and both O and D maps show different areas from which paths are starting (top) or arriving (bottom), that perform rather poorly (yellow $\sim0.5$ and red $\sim0.3$). Paths originating along the Thames (especially along the south bank) seem to be the most impacted in this regime operating below $30\%$ of their peak; other areas are still mostly functional. The situation about paths ending south of the Thames, on the other hand, already shows very poor $P<0.3$. Increasing traffic just by $5\%$ to reach the transition, we observe a slight improvement in $P$ for the origins, especially the northern area, while the south bank of the Thames gets even worse. The D map at the flex seems to improve overall, but this is due to several paths not reaching their destination ($C<1$), but with a good $S$ along $p^*$, so $P$ improves a little overall. Adding $25\%$ more traffic to the city produces a catastrophic result (a state of the network in which most paths cannot get to their destination without using at least a dysfunctional edge) on both maps with $P<0.3$ almost everywhere. 
These $P$ maps show mainly which parts of the network tend to break first, as already seen in previous works~\cite{cogoni2021stability,cogoni2024predicting}, but also how certain areas perform differently when acting as traffic sources or sinks: Some places remain easily reachable during congestion, but, at the same time, leaving them to reach any other destination can be very hard. This asymmetry is partly related to what happens in~\cref{fig:norm_inness} where CW and CCW paths at the same distance from the center could have very different $d$ and $I$, especially at the transition. The maps associated to the other cities are shown in figs.~S37-S44.

\begin{figure}[htb]
    \includegraphics[width=\linewidth]{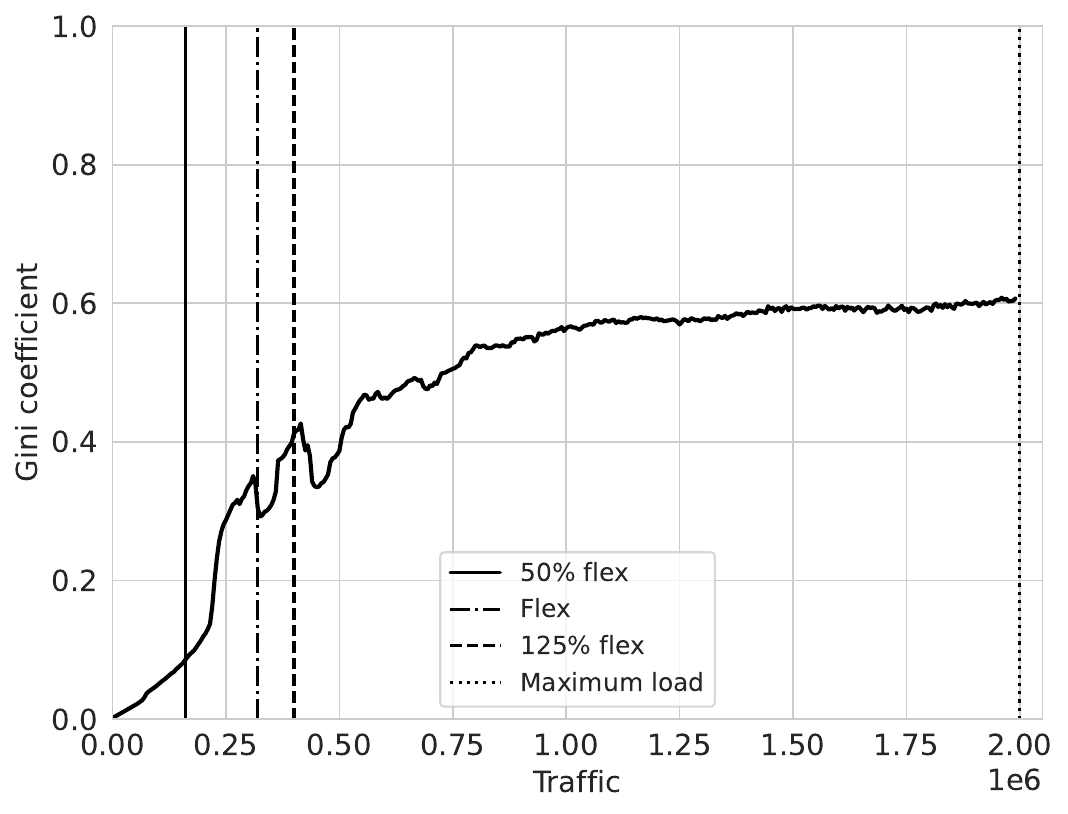}
    \caption{\label{fig:london_gini_coeff}The Gini coefficient as traffic load increases for the city of London. The same traffic levels of~\cref{fig:london_p_map} with the same line styles are plot as vertical lines for comparison.}
\end{figure}

\subsection{Path performance inequality}
Studying path performance from a socio-economical perspective, we could ask ourselves how the network degradation impacts the whole population of paths as congestion grows. It is interesting to study whether paths are uniformly affected or there is an imbalance leading to some paths still being traversable efficiently while the majority is catastrophically broken. Of course, this question makes particular sense near the transition where we know that the system's susceptibility to traffic changes is maximum. In~\cref{fig:london_pi_distr} (see figs.~S45-S46 for the other cities) we see some specific $P$ distributions (four vertical slices from~\cref{fig:london_s_c_p}(bottom)) where we can better observe how $P$ shifts from an average high performance (solid line) at $50\%$ of the transition, which then flattens at the flex and immediately beyond (dot-dashed and dashed lines). 
In \cref{fig:london_gini_coeff} (see figs.~S47-S48 for the other cities), we plot the Gini coefficient at all traffic levels (vertical lines are placed at the same reference levels as in~\cref{fig:london_pi_distr}): at half the flex level the degradation is still moderate ($G\sim0.1$) and balanced (the black solid line in~\cref{fig:london_pi_distr} is mostly localized around $P\sim0.8$) while at the flex (dash-dotted) the distribution flattens and spans the whole $P$ range and the Gini coefficient grows to $\sim0.4$. For even higher traffic we see a $P$ distribution peaked on extremely high path degradation and a Gini over $0.5$, meaning that very few paths are still able to perform well and the rest is almost unusable. The Gini coefficient acts also as a precursor for detecting the transition complementing the soft rejection ratio that we used in~\cref{fig:path_reject_ratio}.

\section{Conclusion}
\label{sec:conclusion}
We have investigated how the FPs on transportation networks evolve as congestion levels increase, using a proven model in which speed decreases linearly with density at the edge level due to interactions among selfish agents. In a different setting, this model has shown the ability to grasp important structural properties of real urban traffic~\cite{cogoni2024predicting}.

In this work, we significantly extend the analysis, shifting the measuring process from an edge-based to a path-based network performance metric. Studying urban vehicular transportation with the selected model, is particularly important for city analysis since the emergence of personal navigation tools, providing the fastest routes for their users in real-time, has shown to have a major impact on traffic patterns~\cite{serok_unveiling_2019}.
We studied the real urban networks of eight major cities, and show the emergence of complex and spatially non-uniform degradation of network performance under growing traffic. In order to quantify this degradation, we introduced a novel measure called Performance Index, which incorporates path slowdown and completeness.

We initially analyzed the low-density network regime, slowly increasing traffic from the empty network, in which a smooth path performance degradation is observed up to a critical traffic volume. Beyond this traffic level, the network abruptly breaks up, as known, into multiple disjoint clusters. Our analysis uncovers the details of the catastrophic decrease of network performance driven by the failure of a very small number of edges: Some cities have a single, clear transition, while others experience a complex sequence of local breakups. By focusing on the dynamically evolving FPs permitted over the network while traffic grows, we show that path shape and performance is a precursor of a change in the global transportation efficiency.
To fully understand the interplay between traffic growth and network efficiency, we characterize the effective length, detour, and inness for the FPs, and study the performance degradation distribution across uniformly distributed paths:
FPs for all cities show a peak for detour and inness (and their variance) in the proximity of the critical traffic level. We discovered that a unique localization of the transition to congestion can be attained by means of multiple performance indicators, but it is easiest to look at the flex of the hard-rejected path ratio curve. For most urban areas, inness uncovers, with light traffic, that paths are slightly attracted by city centers. On the other hand, this central force becomes strongly repulsive during congestion.

By exploiting the proposed path performance index, we demonstrated an important asymmetry in the efficiency of some parts of the cities when operating as traffic sources or as sinks. Finally, the Gini coefficient, applied to path performance, besides being a useful precursor for the congestion transition, quantifies the inequality of degradation over different parts of the network. 

Our model confirmed that, in order to build resilient cities to better exploit their maximum vehicular capacity, naturally involves having multiple available paths among adjacent neighborhoods (that are usually well connected) avoiding single points of failure that tend to saturate early regardless of their capacity.

From the point of view of a city planner, the results present in the literature, relying on real traffic data, do not allow changes to city topology; on the other hand, centrality measures and random percolation do not offer reliable dynamic models to evaluate transportation improvements.
With our method, rapidly iterating over different variants of a city to obtain detailed traffic patterns, potentially allows a deeper insight into the spatial and temporal behavior of the network.

We conclude by observing that the transition to congestion of urban networks involves the existence of one (or more) traffic level(s), that we may define as ``critical'', for which all chosen shape and performance metrics reveal a qualitative change in the networking behavior of all our cities. The transition multiplicity observed for London, Las Vegas, Los Angeles and Paris, stems from inhomogeneous city networks, so that some areas breakup before others (e.g., sparse roads, many one-way streets) or natural barriers (e.g., rivers) limit their connectivity.
This transition to congestion complements the description of the network fragmentation offered by random percolation, through a realistic interaction among agents leading to vastly different edge-usage patterns for low or high density scenarios.

These results promise to offer valuable insights into the vulnerabilities of transportation networks under increasing congestion. Note that, even though in this work our method was used to analyze patterns typical of urban vehicular traffic, it is expected that other transport phenomena involving agent competition for network resources could be approached similarly.

\paragraph*{Acknowledgments --} 
We are deeply indebted to Francesco Versaci for the many useful discussions. We acknowledge the contribution of Sardinian Regional Authorities under project XDATA (art 9 L.R. 20/2015). Map data copyrighted OpenStreetMap contributors: www.openstreetmap.org.


\bibliographystyle{unsrt}
\bibliography{PRE2024}

\begin{thebibliography}{10}

\bibitem{helbing_traffic_2001}
Dirk Helbing.
\newblock Traffic and related self-driven many-particle systems.
\newblock {\em Rev. Mod. Phys.}, 73(4):1067--1141, December 2001.
\newblock Publisher: American Physical Society.

\bibitem{zeng_switch_2019}
Guanwen Zeng, Daqing Li, Shengmin Guo, Liang Gao, Ziyou Gao, H.~Eugene Stanley,
  and Shlomo Havlin.
\newblock Switch between critical percolation modes in city traffic dynamics.
\newblock {\em Proc Natl Acad Sci USA}, 116(1):23--28, January 2019.

\bibitem{zhang2022complex}
Mengyao Zhang, Tao Huang, Zhaoxia Guo, and Zhenggang He.
\newblock Complex-network-based traffic network analysis and dynamics: A
  comprehensive review.
\newblock {\em Physica A: Statistical Mechanics and its Applications},
  607:128063, 2022.

\bibitem{Magnien:2021:IRF}
Cl\'{e}mence Magnien, Matthieu Latapy, and Jean-Loup Guillaume.
\newblock Impact of random failures and attacks on {Poisson} and power-law
  random networks.
\newblock {\em ACM Comput. Surv.}, 43(3), April 2011.

\bibitem{oehlers2021graph}
Milena Oehlers and Benjamin Fabian.
\newblock Graph metrics for network robustness—a survey.
\newblock {\em Mathematics}, 9(8):895, 2021.

\bibitem{schwarze2024structural}
Alice~C Schwarze, Jessica Jiang, Jonny Wray, and Mason~A Porter.
\newblock Structural robustness and vulnerability of networks.
\newblock {\em arXiv preprint arXiv:2409.07498}, 2024.

\bibitem{zeng_multiple_2020}
Guanwen Zeng, Jianxi Gao, Louis Shekhtman, Shengmin Guo, Weifeng Lv, Jianjun
  Wu, Hao Liu, Orr Levy, Daqing Li, Ziyou Gao, H.~Eugene Stanley, and Shlomo
  Havlin.
\newblock Multiple metastable network states in urban traffic.
\newblock {\em Proc Natl Acad Sci USA}, 117(30):17528--17534, July 2020.

\bibitem{cogoni2021stability}
Marco Cogoni and Giovanni Busonera.
\newblock Stability of traffic breakup patterns in urban networks.
\newblock {\em Physical Review E}, 104(1):L012301, 2021.

\bibitem{cogoni2024predicting}
Marco Cogoni and Giovanni Busonera.
\newblock Predicting network congestion by extending betweenness centrality to
  interacting agents.
\newblock {\em Physical Review E}, 109(4):L044302, 2024.

\bibitem{serok_unveiling_2019}
Nimrod Serok, Orr Levy, Shlomo Havlin, and Efrat Blumenfeld-Lieberthal.
\newblock Unveiling the inter-relations between the urban streets network and
  its dynamic traffic flows: {Planning} implication.
\newblock {\em Environment and Planning B: Urban Analytics and City Science},
  46(7):1362--1376, September 2019.
\newblock Publisher: SAGE Publications Ltd STM.

\bibitem{holme2003congestion}
Petter Holme.
\newblock Congestion and centrality in traffic flow on complex networks.
\newblock {\em Advances in Complex Systems}, 6(02):163--176, 2003.

\bibitem{guimera2005worldwide}
Roger Guimera, Stefano Mossa, Adrian Turtschi, and LA~Nunes Amaral.
\newblock The worldwide air transportation network: Anomalous centrality,
  community structure, and cities' global roles.
\newblock {\em Proceedings of the National Academy of Sciences},
  102(22):7794--7799, 2005.

\bibitem{kazerani2009can}
Aisan Kazerani and Stephan Winter.
\newblock Can betweenness centrality explain traffic flow.
\newblock In {\em 12th AGILE international conference on geographic information
  science}, pages 1--9, 2009.

\bibitem{yeung2012competition}
Chi~Ho Yeung and David Saad.
\newblock Competition for shortest paths on sparse graphs.
\newblock {\em Physical review letters}, 108(20):208701, 2012.

\bibitem{hamedmoghadam_percolation_2021}
Homayoun Hamedmoghadam, Mahdi Jalili, Hai~L. Vu, and Lewi Stone.
\newblock Percolation of heterogeneous flows uncovers the bottlenecks of
  infrastructure networks.
\newblock {\em Nat Commun}, 12(1):1254, February 2021.

\bibitem{olmos_macroscopic_2018}
Luis~E. Olmos, Serdar Çolak, Sajjad Shafiei, Meead Saberi, and Marta~C.
  González.
\newblock Macroscopic dynamics and the collapse of urban traffic.
\newblock {\em Proceedings of the National Academy of Sciences},
  115(50):12654--12661, December 2018.
\newblock Publisher: Proceedings of the National Academy of Sciences.

\bibitem{colak_understanding_2016}
Serdar Çolak, Antonio Lima, and Marta~C. González.
\newblock Understanding congested travel in urban areas.
\newblock {\em Nat Commun}, 7(1):10793, April 2016.

\bibitem{diet_towards_2018}
Alexandre Diet and Marc Barthelemy.
\newblock Towards a classification of planar maps.
\newblock {\em Phys. Rev. E}, 98(6):062304, December 2018.
\newblock Publisher: American Physical Society.

\bibitem{aldous2013true}
David Aldous and Karthik Ganesan.
\newblock True scale-invariant random spatial networks.
\newblock {\em Proceedings of the National Academy of Sciences},
  110(22):8782--8785, 2013.

\bibitem{li_percolation_2015}
Daqing Li, Bowen Fu, Yunpeng Wang, Guangquan Lu, Yehiel Berezin, H.~Eugene
  Stanley, and Shlomo Havlin.
\newblock Percolation transition in dynamical traffic network with evolving
  critical bottlenecks.
\newblock {\em Proc Natl Acad Sci USA}, 112(3):669--672, January 2015.

\bibitem{kirkley_betweenness_2018}
Alec Kirkley, Hugo Barbosa, Marc Barthelemy, and Gourab Ghoshal.
\newblock From the betweenness centrality in street networks to structural
  invariants in random planar graphs.
\newblock {\em Nat Commun}, 9(1):2501, June 2018.
\newblock Number: 1 Publisher: Nature Publishing Group.

\bibitem{lee_morphology_2017}
Minjin Lee, Hugo Barbosa, Hyejin Youn, Petter Holme, and Gourab Ghoshal.
\newblock Morphology of travel routes and the organization of cities.
\newblock {\em Nat Commun}, 8(1):2229, December 2017.
\newblock Number: 1 Publisher: Nature Publishing Group.

\bibitem{rakha2002comparison}
Hesham Rakha and Brent Crowther.
\newblock Comparison of greenshields, pipes, and van aerde car-following and
  traffic stream models.
\newblock {\em Transportation Research Record}, 1802(1):248--262, 2002.

\bibitem{ding2018detecting}
Rui Ding, Norsidah Ujang, Hussain bin Hamid, Mohd~Shahrudin Abd~Manan, Yuou He,
  Rong Li, and Jianjun Wu.
\newblock Detecting the urban traffic network structure dynamics through the
  growth and analysis of multi-layer networks.
\newblock {\em Physica A: Statistical Mechanics and its Applications},
  503:800--817, 2018.

\bibitem{kartun2019shape}
Alexander~P Kartun-Giles, Marc Barthelemy, and Carl~P Dettmann.
\newblock Shape of shortest paths in random spatial networks.
\newblock {\em Physical Review E}, 100(3):032315, 2019.

\bibitem{cogoni2017ultrametricity}
Marco Cogoni, Giovanni Busonera, and Gianluigi Zanetti.
\newblock Ultrametricity of optimal transport substates for multiple
  interacting paths over a square lattice network.
\newblock {\em Physical Review E}, 95(3):030108, 2017.

\end{thebibliography}

\end{document}